\documentclass[aps,amsmath,showpacs,twocolumn]{revtex4}
\usepackage{graphicx}

\begin{document}
\title{Cluster algorithm for non-additive 
hard-core mixtures}
\author{Arnaud Buhot}
\affiliation{UMR 5819 (UJF, CNRS, CEA) DRFMC/SI3M, 
CEA Grenoble, 17 rue des Martyrs, 
38054 Grenoble cedex 9, France}

\begin{abstract}
In this paper, we present a cluster algorithm for 
the numerical simulations of non-additive hard-core 
mixtures. This algorithm allows one to simulate 
and equilibrate systems with a number of particles 
two orders of magnitude larger than previous simulations. 
The phase separation for symmetric binary mixtures 
is studied for different non-additvities as well as
for the Widom-Rowlinson model (B. Widom and J. S. Rowlinson,
J. Chem. Phys. 52, 1670 (1970)) in two and three 
dimensions. The critical densities are determined 
from finite size scaling. The critical exponents 
for all the non-additivities are consistent
with the Ising universality class. 
\end{abstract}


\maketitle

\section{Introduction}

The Widom-Rowlinson (WR) model~\cite{Widom} attracted 
a lot of attention as a prototype for the liquid-vapor 
phase separation. This simple model is composed of a 
two-component system where likewise particles do not 
interact whereas unlike particles interact through a 
hard-core potential. It was shown to present a phase 
separation at high density and a critical point that 
belongs to the Ising universality class~\cite{Shew,Johnson}. 
This mixture exhibits a liquid-liquid critical point 
(with large composition fluctuations) in contrast
to pure fluids that experience a liquid-vapor critical 
point (with large density fluctuations). However, the
phase transitions are related~\cite{Hohenberg} and
believed to depend on the same universality 
class~\cite{Jagannathan2}.

A straightforward generalization of the 
WR model is the non-additive 
hard-core (NAHC) mixtures where the 
likewise particles also experience a 
hard-core interaction. Two particles $i$ 
and $j$ present a minimal distance of 
approach $\sigma_{XY}$ with $X$ and $Y$ 
representing respectively the $A$ or $B$ 
component from which belong particles 
$i$ and $j$. Non-additive mixtures are 
characterized by $\sigma_{AB} = (\sigma_{AA}
+\sigma_{BB})(1+\Delta)/2$ with $\Delta \neq 
0$. Additive mixtures correspond to $\Delta 
= 0$. For a negative non-additivity $\Delta 
< 0$, particles tend to form 
hetero-coordinations~\cite{Gazzillo,Gazzillo2}.
For a positive non-additivity $\Delta > 0$, an 
entropically driven phase separation occurs 
between two phases at sufficiently high density 
due to the extra repulsion between unlike 
particles~\cite{Frenkel}. The two phases are 
chemically different, one is rich in $A$ particles 
whereas the other is rich in $B$ particles.
This phase separation occurs even for the symmetric 
mixtures where $\sigma_{AA} = \sigma_{BB}$. 
The critical density as well as the universality 
class have been determined from different numerical 
simulations~\cite{Saija,Saija2,Jagannathan,Gozdz,Jagannathan2}.
The special case of $\sigma_{AA} = \sigma_{BB} = 0$ 
corresponding to the WR model has also 
been studied~\cite{Shew,Johnson}.
Additive mixtures with a strong asymmetry
$\sigma_{AA} \ll \sigma_{BB}$ are also of interest 
and present another kind of entropically driven phase 
separation transition predicted for the first time by
Biben and Hansen~\cite{Biben} but only recently observed
by numerical simulations~\cite{Buhot,Dijkstra}. 

Most recent simulations of the NAHC 
mixtures~\cite{Jagannathan,Gozdz,Jagannathan2} used 
a Monte-Carlo algorithm within the semigrand 
canonical ensemble~\cite{Kofke,Miguel}.
In addition to the simple moves of particles 
as simple Monte-Carlo steps, some steps consist 
in changing the nature (or component) of a particle
when the hard-core interactions permit this 
modification. The resulting ensemble corresponds 
to a fixed total number of particles but a variable 
composition (or fixed difference of chemical potentials 
between particles of different components). 
A detailed description of the algorithm in the 
case of binary mixtures with squared-well interactions 
may be found in the paper of de Miguel 
{\it et al.}~\cite{Miguel}. 
Working in the semigrand canonical ensemble gives 
access to the coexistence curve of the model.  
However, the algorithm is limited to rather small 
system sizes. The largest simulations concerned 
$16 384$ particles~\cite{Jagannathan} but most of them 
were limited to a few thousands~\cite{Saija,Saija2,Gozdz}. 
Furthermore, only small non-additivities $\Delta \leq 1$ 
have been simulated~\cite{Gozdz,Jagannathan}. The main 
reason for this limitation is the following. When a 
Monte-Carlo step corresponding to a change of components 
is tried, at high non-additivity there is a large 
probability of overlap with at least one of the neighbors. 
This results in the rejection of the change of components 
and a dramatic slowing down of the equilibration of the 
numerical simulations. 

In the present paper we consider a cluster 
algorithm proposed by Dress and Krauth~\cite{Dress} 
for hard-core mixtures which proved useful for the 
detection of the phase separation in additive 
asymmetric mixtures~\cite{Buhot,Buhot2,Buhot3} and for 
the analysis of two dimensional polydisperse hard-core 
mixtures~\cite{Santen}. The advantage of this 
algorithm is to allow one to equilibrate systems 
with up to $10^6$ particles (two order of magnitude 
larger than previous simulations). Also, the slowing 
down of the equilibration for large non-additivity 
is completely avoided with this cluster algorithm. 
This allows us to analyze the limit of infinite 
non-additivity ($\Delta \rightarrow \infty$) where 
the NAHC mixtures converge to the WR 
model. Furthermore, the coexistence curve is accessible 
in contrary to the previous cluster algorithm used 
for the WR model~\cite{Johnson}. 

The paper is organized as follows: in section II, 
we present the model of non-additive hard-core 
mixtures and we describe the cluster algorithm 
used for the numerical simulations. We present the 
numerical results for the two and three dimensional
mixtures in section III. From finite size 
scaling analysis, we extract the critical densities 
and the critical exponents. Finally, we conclude 
by a discussion of the results in section IV. 
 
\section{Non-Additive Hard-Core Mixtures}

\subsection{Description of the model} 

We consider non-additive 
hard-core (NAHC) mixtures in two and three 
dimensions. The system is made of two 
components $A$ and $B$ with respectively 
$N_A$ and $N_B$ particles. The particles 
experience hard-core interactions. No overlap 
is possible if the distance $d$ between the 
center of the particles $i$ and $j$ is 
lower than $\sigma_{XY}$ where $X$ and $Y$ 
represent the components of particles $i$ 
and $j$ respectively. For convenience, the
particles are placed in two identical boxes 
of equal volume $V$. Periodic boundary 
conditions are assumed on each of the boxes. 
From the following description of the cluster 
algorithm, the reason for considering two boxes 
will become obvious. Notice that the consideration 
of two boxes is of common use for the determination 
of the coexistence curve in semigrand canonical 
ensemble simulations.

In the following, we restrict ourselves to 
symmetric mixtures with $\sigma_{AA} = \sigma_{BB} 
= \sigma$ however we allow for a general positive
non-additivity $\Delta = \sigma_{AB}/\sigma -1$.
Due to the hard-core interactions, the temperature 
plays a trivial role and the phase diagram is 
determined only by the number density $\rho = 
\rho_A + \rho_B = N_A/2V + N_B/2V$ and the 
composition $x_A = 1 - x_B = \rho_A/(\rho_A + 
\rho_B$) of the system. In the particular case 
of the symmetric NAHC mixtures, the critical 
point ($\rho_c,x_c$) is determined in composition 
($x_c = 1/2$) due to the symmetry. Thus, in the 
following, we will consider $N_A = N_B$. 
Above the critical density $\rho_c$, the system 
separates in two phases $I$ and $II$. One phase is 
rich in $A$ particles and the other is rich in $B$ 
particles. Furthermore, due to the symmetry, 
those phases are symmetric in composition 
($x_A^I = x_B^{II}$ and $x_A^{II} = x_B^I$). 

The determination of the coexistence curve $x_A^{I}$
and $x_A^{II}$ is possible from the use of the two 
equivalent boxes. The overall composition $x_A = 
N_A/(N_A+N_B)$ is fixed during the simulations 
whereas the particular compositions $x_A^I = 
N_A^I/(N_A^I+N_B^I)$ and $x_A^{II} = 
N_A^{II}/(N_A^{II}+N_B^{II})$ of the boxes 
$I$ and $II$ are free to fluctuate. $N_X^{Y}$ 
corresponds to the number of particles of component 
$X = A$ or $B$ inside the box $Y = I$ or $II$. 
Notice that the number of particles in each box 
$N^{I} = N_A^{I} + N_B^{I}$ and $N^{II} = N_A^{II}+ 
N_B^{II}$ also fluctuates. The ensemble considered 
is thus intermediate between the grand canonical 
ensemble and the semigrand canonical ensemble. 
We will discuss in the following the consequences 
of the slight density fluctuations inside each box. 

In complement to the number density $\rho$, it is 
interesting to introduce the scaled packing fraction 
$\eta = v_{AB} \rho$ with $v_{AB}$ the volume of 
unlike particles. $v_{AB} = \pi \sigma_{AB}^2/4$ in 
two dimensions and $\pi \sigma_{AB}^3/6$ in three 
dimensions. This definition of the packing fraction 
which takes into account the non-additivity allows 
us to compare the critical packing fraction of the
WR model with those of the NAHC 
mixtures. Notice that from this definition, the 
packing fraction may exceed one. 

\subsection{Description of the cluster algorithm}

A cluster algorithm has been introduced recently 
by Dress and Krauth~\cite{Dress} for the simulation 
of hard-core mixtures. Inspired by the lattice cluster
algorithms of Swendsen-Wang~\cite{Swendson} and 
Wolff~\cite{Wolff}, it allows one to equilibrate large 
off-lattice systems (up to $10^6$ particles) thanks to 
a non-local move of a large number of particles at each 
Monte-Carlo step. The general idea of the algorithm is 
to take advantage of the hard-core interactions between 
particles to construct a cluster of particles that will 
be moved at each Monte-Carlo step satisfying the detailed 
balance while keeping ergodicity. 

The Monte-Carlo step is constructed as follows: 

i) We select randomly one of the two boxes. 

ii) A second box is chosen randomly (it can be 
the same as the previous one). 

iii) An inversion symmetry around a randomly chosen 
pivot point is performed on all the particles of the 
second box. 

iv) The two boxes with their particles are then 
superimposed on top of each other resulting in a 
set of clusters of overlapping particles (in the 
sense of the hard-core interactions). 

v) A particle is randomly  selected and the cluster 
from which it belongs is flipped. Each particle 
belonging to this cluster is moved from its initial 
box to the other box in the position corresponding 
to the superimposed configuration. 

For a clear graphical representation of a Monte-Carlo 
step see Figure 1 in Dress and Krauth paper~\cite{Dress}.

Let us first prove that the new configuration obtained
after the Monte-Carlo step satisfies all the hard-core 
interactions between particles. a) For two particles 
outside the flipped cluster there is no overlap since 
those particles did not move and there was no overlap 
before the Monte-Carlo step. b) For two particles 
inside the cluster, they were flipped keeping there 
relative positions in such a way that there is still 
no overlap. c) Finally, by construction of the 
cluster, there is no overlap between a particle 
belonging to the cluster and one outside the cluster 
when the two boxes are superimposed and {\it a 
fortiori} after the Monte-Carlo step. Those arguments 
justify the fact that the new configuration obtained 
satisfies all the hard-core interactions between 
particles in each box. 

Now let us justify the detailed balance of the 
cluster algorithm. Each Monte-Carlo step has 
its symmetric step in the sense that if we start 
from the final configuration there is a pivot 
point that gives back the initial configuration. 
The pivot points being randomly selected with 
a uniform distribution, there are equality between 
the probabilities to go from the initial to the 
final configuration and vice-versa. The detailed 
balance is thus satisfied noting that due to the 
hard-core interactions both configurations have 
the same equilibrium probability. In the case of 
a general potential of interaction between particles, 
the cluster algorithm may be generalized with the 
extent of some rejections of the Monte-Carlo 
steps~\cite{Malherbe}.

On the point ii) of a Monte-Carlo step, either 
the same box or the other one may be selected. 
This leads to intra-box or inter-box 
Monte-Carlo steps. The reason for this choice 
is to decrease the equilibration time and as 
we will see to satisfy the ergodicity of the 
cluster algorithm. With intra-box Monte-Carlo 
steps (same box selected twice), it has been 
shown already~\cite{Dress,Buhot} that the 
algorithm satisfies internal ergodicity like 
the usual Monte-Carlo algorithm. By internal 
ergodicity, we consider the fact that for the 
given composition of the box, all possible 
configurations of the particles are attainable 
by the cluster algorithm. In fact, if the pivot 
point is chosen sufficiently close to a particle 
position and this particle considered as the 
starting point for the cluster construction, 
the Monte-Carlo step corresponds to a slight 
move of this particle without affecting the 
other particles. This move corresponds to a 
usual Monte-Carlo move in a general algorithm. 
This argument justifies the internal ergodicity.
However, from those moves, the composition of 
the box is kept constant. In order to change the
composition or the relative number of particles 
of components $A$ or $B$ in each box, inter-box 
Monte-Carlo steps are necessary. Those inter-box 
Monte-Carlo steps are performed to exchange the 
components of particles. If the pivot point is 
selected such that two particles from different 
components and boxes superimposed exactly and if 
one of those particles is selected as the cluster 
starting point, the Monte-Carlo step reduces 
to the exchange of those particles. 
The Monte-Carlo move then corresponds to the 
usual change of components considered by the 
Monte-Carlo algorithm in the semigrand 
canonical ensemble. In summary, both usual 
Monte-Carlo steps (the move of a 
particle and the change of components) are 
possible moves in this cluster algorithm 
justifying the ergodicity if this ergodicity 
is assumed for the usual Monte-Carlo 
algorithm in the semigrand canonical ensemble.

As can be seen from the discussion on the 
ergodicity, the use of the two boxes is useful 
for the non-additive hard-core mixtures. 
It has another strong advantage since it
allows one to determine the coexistence curve
or the relative composition of the two 
separated phases above the critical density.
Due to the symmetry of the problem, the 
critical point corresponds to an equal 
partition in particles $A$ and $B$ ($x_c 
= 1/2$) and above the critical density the 
coexistence curve is symmetric around $x_c=1/2$. 
The use of identical boxes is thus justified 
since the densities $\rho^I$ and $\rho^{II}$ 
of the two phases are equal. 
As previously discussed, the total number of 
particles in each box is slightly fluctuating 
with this cluster algorithm in contrary to the
simulations in the semigrand canonical ensemble. 
However, those fluctuations of the density are 
not related to the composition fluctuations 
inside each box. More importantly, around the 
phase separation transition, the composition  
fluctuations diverge whereas the density 
fluctuations stay insensitive to the transition. 
As a consequence the slight density fluctuations 
do not affect the coexistence curve determined 
from the cluster algorithm.

The last question concerning the cluster 
algorithm concerns the equilibration time 
or the number of Monte-Carlo steps necessary 
to equilibrate the system. The Swendson-Wang 
cluster algorithm was introduced to simulate
systems of Ising spins with ferromagnetic 
interactions between neighbor spins on a 
lattice~\cite{Swendson}. In contrary to the 
simple Monte-Carlo algorithm, this cluster 
algorithm does not suffer from a critical 
slowing down at the phase transition due to 
the fact that the flipped clusters are then 
directly related to the spin clusters observed 
around the phase transition. In the case of the 
present cluster algorithm, such direct relation 
of the flipped clusters with the configurations 
of particles is not demonstrated. However, 
the number of Monte-Carlo steps necessary for
the equilibration of the system does not seem 
to increase significantly when approaching the 
phase separation transition. It is also interesting 
to note that this number is roughly independent of 
the system size as observed for cluster algorithms 
on lattices and in contrary to usual Monte-Carlo 
algorithms where this number usually increase 
strongly with the system size. 
Another important point is that 
there is no critical slowing down for the 
equilibration when the non-additivity is 
increased. Thus, this cluster algorithm is well 
adapted for large non-additivities $\Delta$ and 
for the WR model in comparison to
the Monte Carlo simulations in the semigrand
canonical ensemble. In fact, the critical slowing
down is observed for small non-addtivities when
the phase separation transition occurs at high 
densities close to a fluid-solid transition. 

\section{Results of the numerical simulations}

\subsection{Critical exponents and critical 
packing fractions}

\begin{table}
\begin{tabular}{|c|c|c|c|}
\hline
Exponent & $\nu$ & $\gamma$ & $\beta$ \\
\hline
\ 2D (exact) \ & 1 & 7/4 & 1/8 \\
\hline
3D (num.)& \ $0.627$ \ & \ $1.239$ \ & 
\ $0.326$\ \\
\hline
\end{tabular}
\caption{\label{exponents}
Critical exponents for the Ising universality 
class in two and three dimensions.}
\end{table}

As all second order phase transitions, the critical 
point in NAHC mixtures is characterized by different 
critical exponents. From finite size scaling of 
equilibrium thermodynamic quantities close to the 
critical point ($\rho_c,x_c$), it is possible to 
determine those exponents. For the NAHC mixtures
and for the WR model, the critical 
exponents are supposed to belong to the Ising 
universality class (see numerical values of $\nu, 
\gamma$ and $\beta$ in Table~\ref{exponents}). 
In the following, we define different ways to 
determine or test the value of the critical 
exponents. We also describe four different ways to 
define a finite size critical packing fraction
$\eta_c(L)$ from numerical simulations. 

From finite size scaling, it is possible to 
extract the (infinite size) critical packing 
fraction $\eta_c$~\cite{Gozdz,Binder}:
\begin{equation}
\label{nu}
\eta_c(L) = \eta_c - A/L^{1/\nu}.
\end{equation}
The finite size of the system $L$ is defined 
as $N_A = L^d$ where $d$ is the dimension of 
the system. The critical exponent $\nu$ is 
independent of the definition of the finite 
size critical packing fraction considered in 
contrary to the coefficient $A$. Due to the 
corrections to scaling for small system sizes,
it is usually difficult to extract the 
critical exponent $\nu$. However, a linear 
behavior of the finite size critical packing 
fraction with the rescaled system size 
$1/L^{1/\nu}$ is a good test of the 
universality class of the model considered and
allows one to extract the critical packing
fraction $\eta_c$.

The phase separation transition is characterised
by the order parameter $m = 2 (x_A - x_c)$ with
$-1 \leq m \leq 1$. Its probability distribution 
$P(m;\eta,L)$ for a packing fraction $\eta$ and a 
system size $L$ changes form around the critical 
packing fraction $\eta_c$. In the thermodynamic 
limit ($L = \infty$), the distribution of the 
order parameter presents delta picks. It has a 
single pick at $m = 0$ below the critical packing 
fraction ($\eta < \eta_c$). However, above the 
critical packing fraction ($\eta > \eta_c$), the 
distribution is double picked at values $\pm 
m_{\max}$ with $m_{\max} \sim (\eta-\eta_c)^{\beta}$ 
for $\eta \gtrsim \eta_c$. For finite size systems, 
the picks broaden but the change of the distribution 
remains and a maximum $m_{\max}(\eta,L)>0$ may be 
defined for each packing fraction $\eta$ and size 
$L$ above the size dependent critical packing 
fraction $\eta_c^{\max}(L)$. The dependence of the 
maximum $m_{\max}$ on the packing fraction at a 
fixed system size is~\cite{Binder}:
\begin{equation}
m_{\max} \sim (\eta - \eta_c^{\max}(L))^{\beta}
\end{equation}
\noindent for $\eta \gtrsim \eta_c^{\max}(L)$. 
Knowing the exponent $\beta$, it is thus possible 
to extract the finite size critical packing 
fraction $\eta_c^{\max}(L)$ from a linear fit 
of $m_{\max}^{1/\beta}$ as function of $\eta$. 
For high packing fraction $\eta$, $m_{\max}$ 
saturates to one. This limits the range of 
packing fractions for which the linear fit is 
valid. Due to this limited range of the power 
law behavior, it is difficult to extract 
the critical exponent $\beta$. However, a 
linear dependence obtained for the expected 
value of the critical exponent $\beta$ is 
still a strong confirmation of the 
universality class.

Due to the symmetry of the model, the 
average of the order parameter is zero. 
However, the mean absolute value of the 
order parameter is another possibility to 
define a finite size order parameter:
\begin{equation}
\langle | m | \rangle(\eta,L) = \int_{-1}^{1} 
|m| P(m;\eta,L) dm.
\end{equation}
This definition suffers from an additional 
drawback compared to $m_{\max}$.  The average 
absolute value of the order parameter does 
not vanish at the finite size critical packing 
fraction $\eta_c^{av}(L)$. Thus, the power law 
behavior~\cite{Binder}:
\begin{equation}
\label{power}
\langle | m | \rangle \sim (\eta - \eta_c^{av}(L))^{\beta}
\end{equation}
presents corrections to scaling not only for large 
packing fraction but also around $\eta_c^{av}(L)$. 
This finite size critical packing fraction 
$\eta_c^{av}(L)$ may still be extracted from a 
linear fit of $\langle |m| \rangle^{1/\beta}$ on 
a limited range of packing fractions. 
The small value of the critical exponent $\beta$ 
in two dimensions leads to a sufficiently large range, 
however, in three dimensions, the larger value of 
$\beta$ renders the range of power law behavior 
(\ref{power}) too small to be able to extract 
$\eta_c^{av}(L)$. 

\begin{table}
\begin{tabular}{|c|c|c|c|c|c|}
\hline
$\Delta$ & $0.5$ & $1.0$ & $2.0$ & $\infty$ & \ Ising \ \\
\hline
\ 2D \ & \ $1.749(8)$ \ & \ $1.749(8)$ \ & \ 
$1.746(7)$ \ & \ $1.742(8)$ \ & $7/4$ \\
\hline
\ 3D \ & $2.03(8)$ & $2.05(8)$ & $2.05(8)$ & $2.05(8)$ 
& \ $1.98$ \ \\
\hline
\end{tabular}
\caption{\label{gammanu}
Numerical results for the ratio of critical exponents 
$\gamma/\nu$ for different non-additivities and for the
WR model ($\Delta = \infty$) in two and 
three dimensions. Numbers in parenthesis correspond to 
the error on the last digits. The data in column Ising 
are the predictions of the Ising universality class.}
\end{table}

The maximum of the modified susceptibility $\chi(\eta,L) 
= \langle m^2 \rangle - \langle |m| \rangle^2$ is a 
third possibility to define a finite size critical 
packing fraction. This modified susceptibility presents 
a single maximum in contrary to the real susceptibility 
$\langle m^2 \rangle - \langle m \rangle^2$. The general 
form of the modified susceptibility is~\cite{Binder}:
\begin{equation}
\chi(\eta,L) = L^{\gamma/\nu-d} \tilde{\chi}
(L^{1/\nu} (\eta-\eta_c))
\end{equation}
with $\tilde{\chi}(x)$ a function with a single 
maximum $\tilde{\chi}_{\max}$. The packing fraction 
at the maximum of the modified susceptibility 
defines the finite size critical packing fraction 
$\eta_c^{\chi}(L)$. The modified susceptibility may 
also be used to extract the ratio of critical 
exponents $\gamma/\nu$. Its maximum depends 
algebraically on the system size with an exponent 
$\gamma/\nu - d$~\cite{Binder}: 
\begin{equation}
\chi_{\max}(\eta_c^{\chi}(L),L) \sim 
L^{\gamma/\nu-d} \tilde{\chi}_{\max}.
\end{equation}
The ratio of critical exponents $\gamma/\nu$ 
is thus simply determined by the slope of a 
linear fit in a log-log scale. The results for 
the two and three dimensional systems are 
presented in Table~\ref{gammanu} in comparison 
with the Ising universality prediction.
Those results are discussed later.  

Another possibility to determine the critical 
packing fraction concerns the Binder 
parameter~\cite{Binder}:
\begin{equation} 
U(\eta,L) = 1 - \frac{\langle m^4 \rangle}{3 
\langle m^2 \rangle}. 
\end{equation}
The Binder parameter saturates to $2/3$ for 
large packing fractions and vanishes for small 
ones. It also presents the interesting 
property to intersect at the critical packing 
fraction at least for sufficiently large system 
sizes. The value $U(\eta_c,L) = U^*$ at this 
intersection is expected to be universal.
The monotonous behavior of the Binder 
parameter may be used to define a finite size 
critical packing fraction $\eta_c^B(L)$ as the 
value for which $U(\eta_c^B(L),L) = 1/2$.
The choice of the value $1/2$ is arbitrary 
and could be modified as soon as it is 
sufficiently different from the boundary values 
$0$ and $2/3$ and from the intersction value $U^*$. 
The particular choice of $U^*$ would render 
the finite size packing fraction independent 
of the system sizes. This point will be 
discussed further for the three dimensional 
systems. 

We defined different critical packing fractions 
for finite size systems from the maximum of the 
modified susceptibility $\eta_c^{\chi}(L)$, from 
the Binder parameter $\eta_c^{B}(L)$ and from the 
coexistence curve either from the maximum of the 
distribution of the order parameter $\eta_c^{\max}
(L)$ or from the average absolute value of the 
order parameter $\eta_c^{av}(L)$. The numerical 
results obtained from the cluster algorithm are 
analyzed in the following sections. 

\subsection{Two dimensional NAHC mixtures}

In the two dimensional model, we considered
four different non-additivities $\Delta = 0.5, 
1.0, 2.0$ and $4.0$ as well as the WR 
model ($\Delta = \infty$). The system sizes ranged 
from $L = N_A^{1/2} = 10$ to $400$. The largest 
systems contained $N_A + N_B = 320 000$ particles.
Numerical simulations were divided in five consecutive 
runs of $10^5$ to $10^6$ Monte-Carlo steps depending 
on the system sizes with longest runs for the largest 
systems. The first run is kept for equilibration 
of the initial configuration and the last four 
runs for the data collection and error estimation. 

\begin{figure}
\includegraphics[width=8cm]{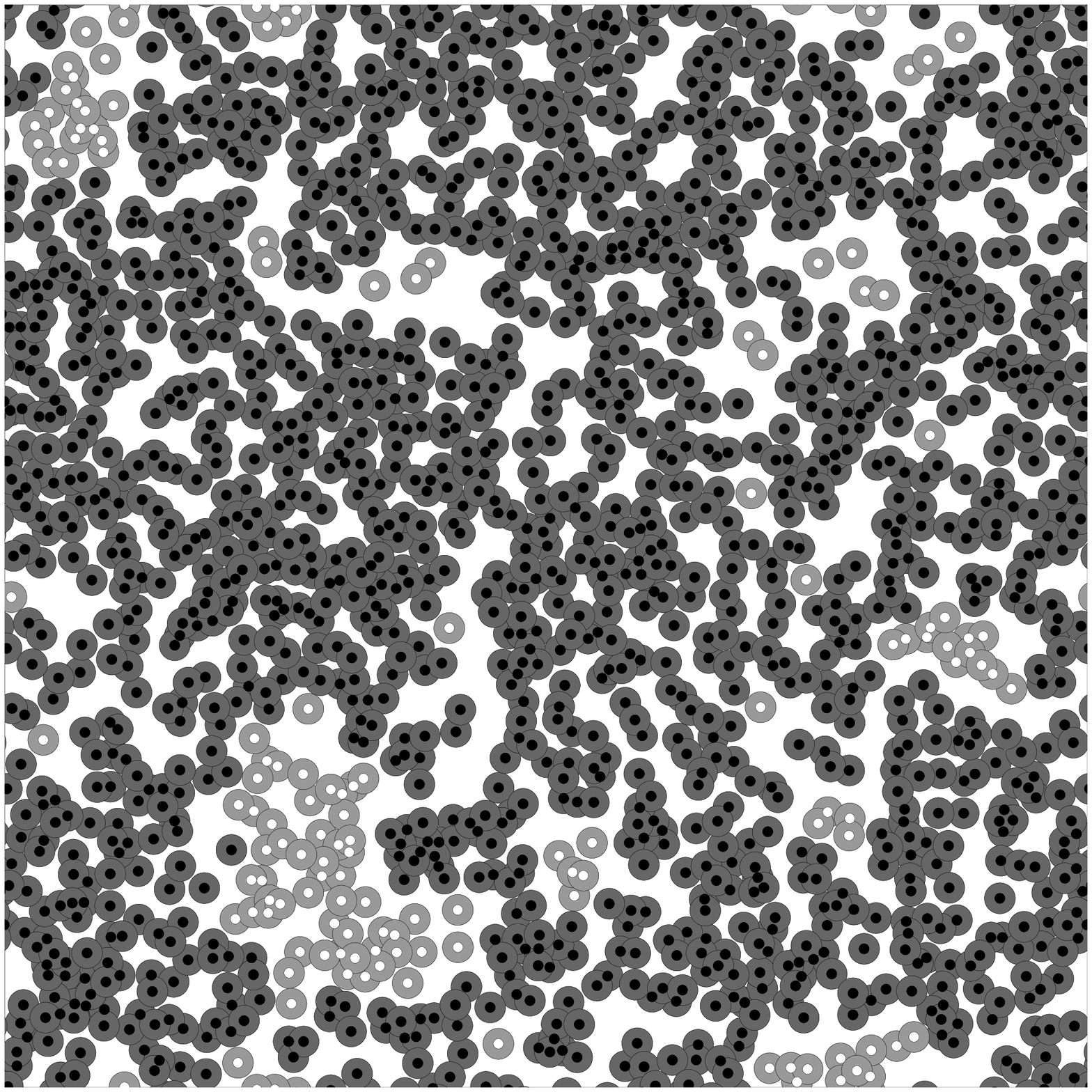}
\includegraphics[width=8cm]{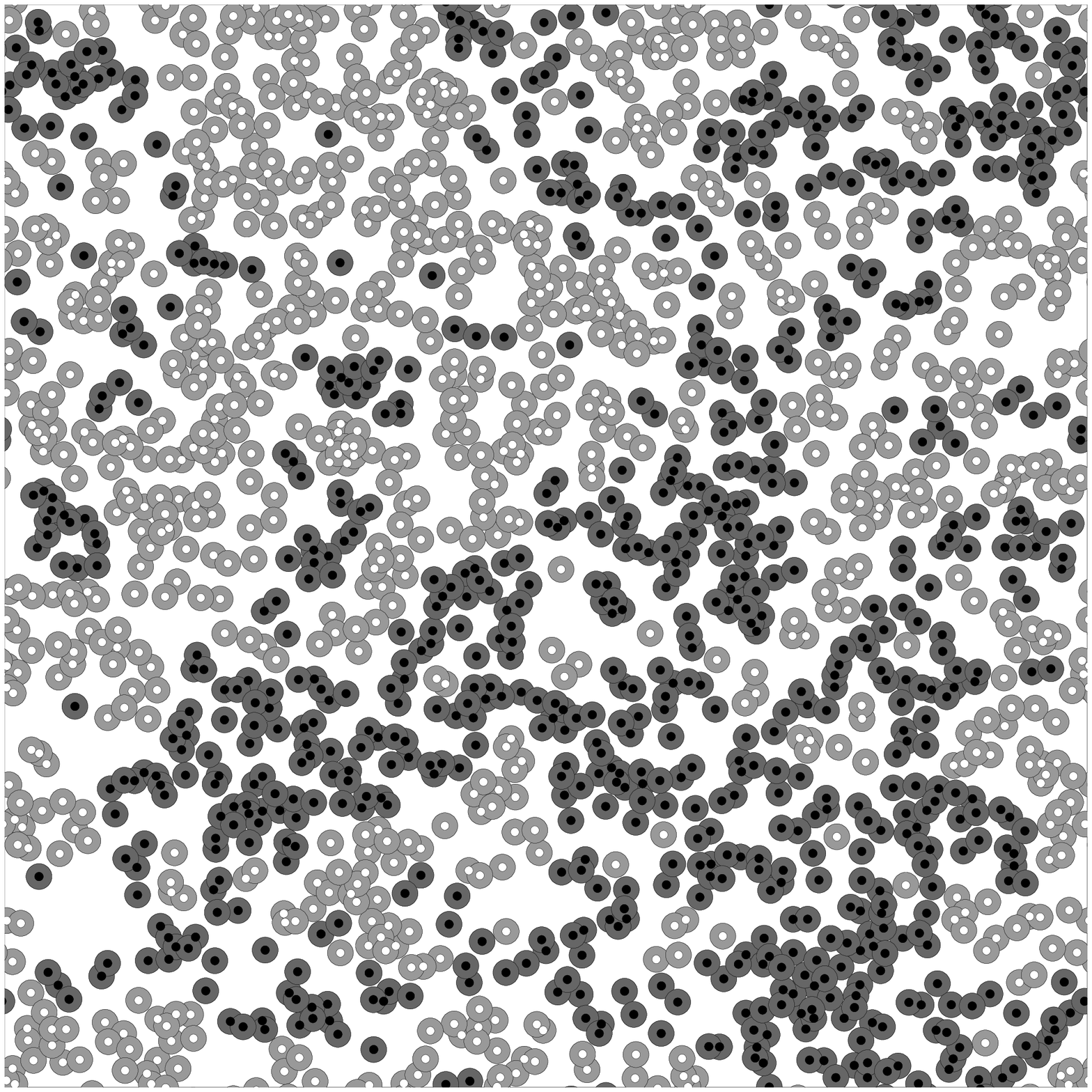}
\caption{\label{config}
Equilibrium configurations for a system above (top)
and below (bottom) the phase separation transition.
The systems correspond to $N_A + N_B = 3200$ particles
with a non-additivity $\Delta = 2$. White and black 
disks correspond to likewise diameter $\sigma_{AA}$ 
and $\sigma_{BB}$ whereas the light and dark grey 
disks correspond to the unlike diameter $\sigma_{AB}$. 
With this representation, the overlaps are only allowed 
by the hard-core interactions to the likewise particles 
on the grey scale disks.}
\end{figure}

On Fig.~\ref{config}, we plot equilibrium
configurations for two different packing fractions.
The configurations correspond to systems with 
$N_A + N_B = 3200$ particles and a non-additivity 
$\Delta = 2$. The likewise hard-core diameter $\sigma_{AA}$ 
and $\sigma_{BB}$ are represented respectively by white 
and black disks whereas the unlike diameter $\sigma_{AB}$ 
is represented by light and dark grey disks respectively
for the $A$ and $B$ particles. No overlaps between 
unlike particles are present but overlaps of likewise 
particles are observed on the grey scale. However, no 
overlap is present between black and white particles 
justifying that both configurations satisfy all the 
hard-core interactions. For the top configuration 
above the phase separation transtion, $\eta = 1.03 > 
\eta_c(L) \simeq 0.95$, we observe a large difference 
between the number of $A$ and $B$ particles. The 
large percolating cluster of $B$ particles (the dark 
grey particles) is a clear evidence of the phase 
separation. On the contrary, on the bottom configuration, 
$\eta = 0.72 < \eta_c(L)$, the $A$ and $B$ particles 
are roughly in equal number and perfectly mixed. 
In this systems, the phase separation did not occur, 
however, due to the stronger unlike particles hard-core 
interactions, a local clustering of likewise particles
is present.

\begin{figure}
\includegraphics[width=8cm]{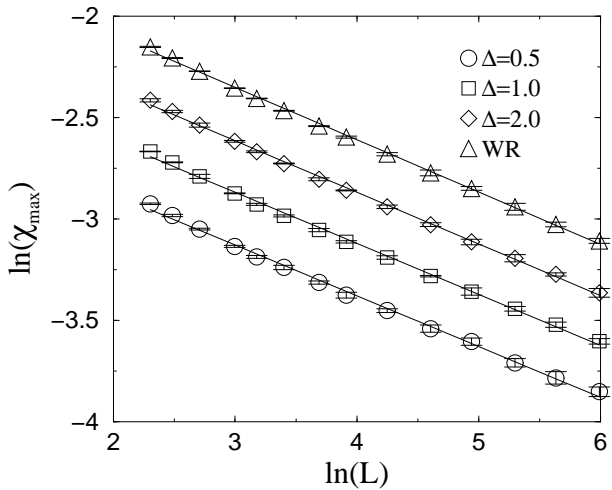}
\caption{Maximum of the modified susceptibility 
$\chi_{\max}$ as function of the system size $L$ 
in a log-log scale for three different 
non-additivities $\Delta$ and for the WR model 
(Widom).  The ratio of critical exponents 
$\gamma/\nu$ is determined from linear fits. 
The three smallest system sizes are removed 
from the linear fit due to the corrections to 
scaling effects for those small sizes. 
The data have been shifted for clarity.
\label{susmax2}}
\end{figure}

The modified susceptibility $\chi$ is determined 
as function of the packing fraction $\eta$ for 
different system sizes. It presents a single maximum 
$\chi_{\max}(\eta_c^{\chi}(L),L)$ at the finite size 
critical packing fraction $\eta_c^{\chi}(L)$. 
As can be seen on Fig.~\ref{susmax2}, this maximum 
depends on the system size with a power law behavior. 
From a linear fit in a log-log scale, we deduce 
the ratio of critical exponents $\gamma/\nu$ 
(see Table~\ref{gammanu}). Due to the corrections 
to scaling observed for small system sizes, 
the three smaller sizes ($L =10, 12$ and $15$) where 
removed for the linear fit to extract $\gamma/\nu$. 
This ratio of critical exponents compare nicely 
with the Ising universality class prediction $7/4$
for all the non-additivities considered and for the 
WR model. The estimation for the errors presented 
on Table~\ref{gammanu} comprises the error on the 
maximum of the modified susceptibility and the error 
coming from the linear fit. The relative error on 
the ratio of critical exponents $\gamma/\nu$ is 
smaller than $1\%$. 

\begin{figure}
\includegraphics[width=7.5cm]{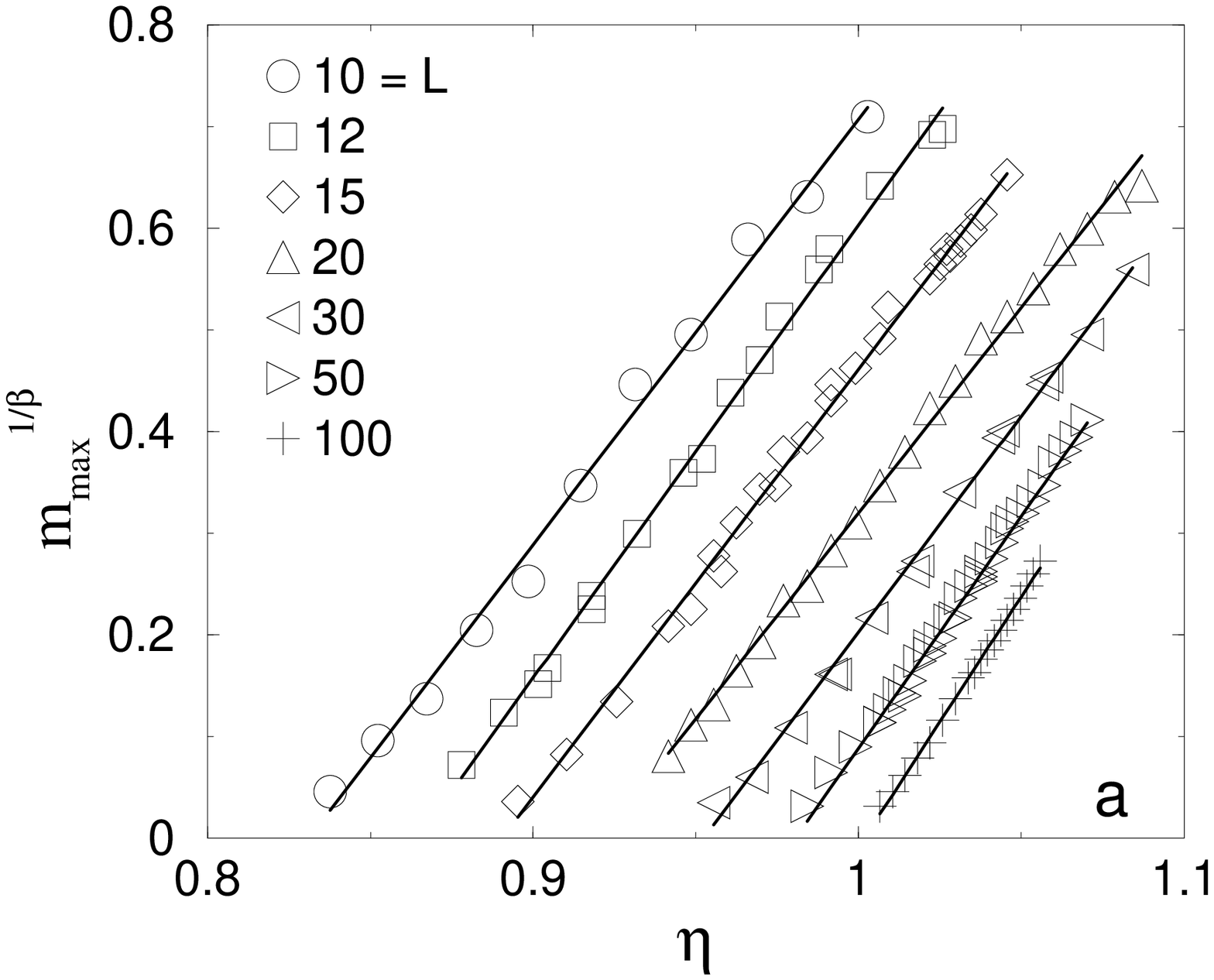}
\includegraphics[width=7.5cm]{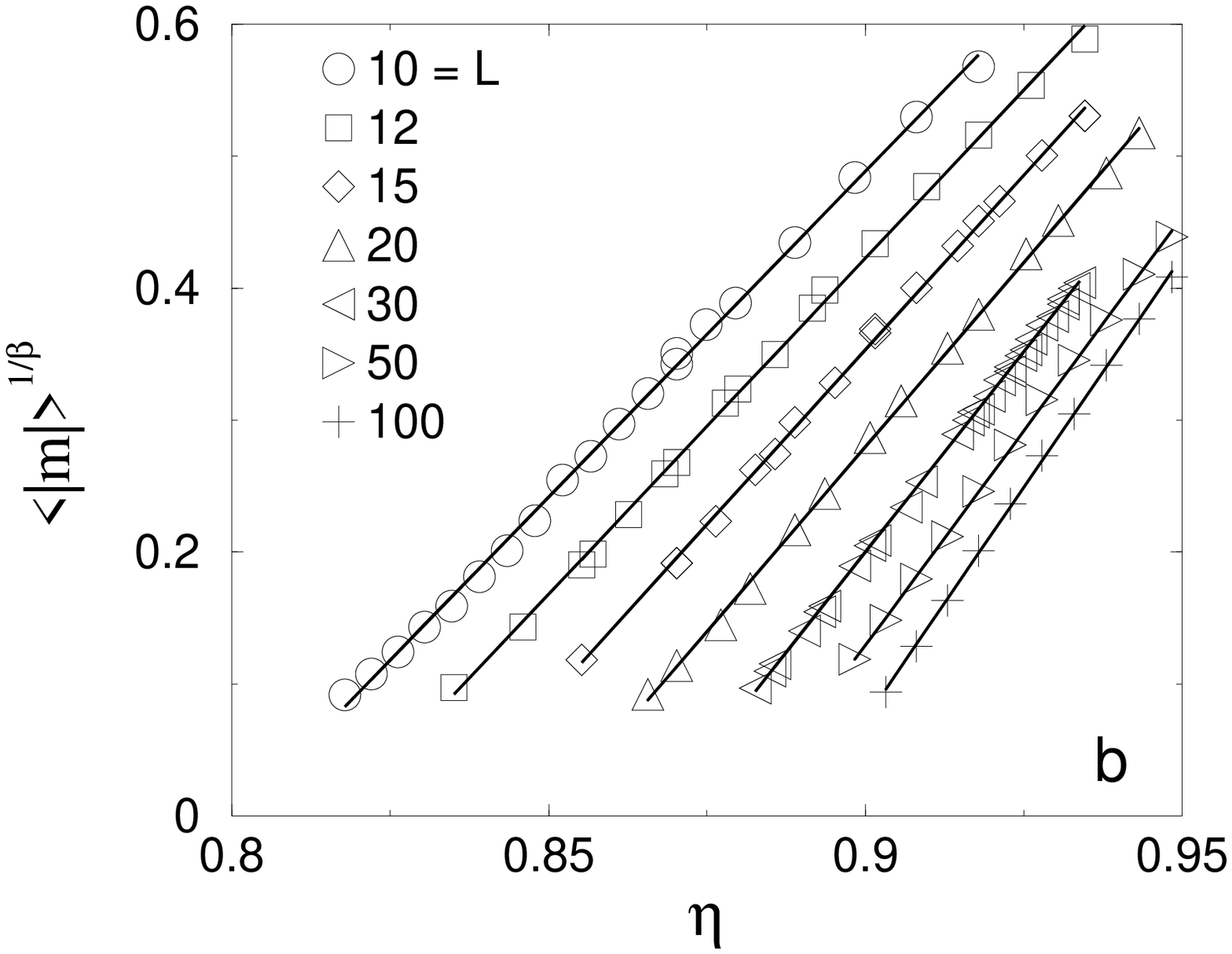}
\caption{\label{mean2}
(a) Rescaled maximum of the distribution 
of the order parameter $m_{\max}^{1/\beta}$ 
for a non-additivity $\Delta = 2$ and
(b) rescaled average order parameter 
$\langle | m | \rangle^{1/\beta}$ for a 
non-additivity $\Delta = 1$ as function of 
the packing fraction $\eta$ for different 
system sizes. Linear fits confirm the critical 
exponent $\beta = 1/8$ predicted by the Ising 
universality class and allow us to extract the 
finite size critical packing fractions 
$\eta_c^{\max}(L)$ and $\eta_c^{av}(L)$ 
respectively.}
\end{figure}

The confirmation of the critical exponent $\beta$
is obtained from the two different definitions of
the finite size order parameter: $m_{\max}$ and
$\langle |m| \rangle$. First, we plot the rescaled 
maximum of the distribution of the order parameter 
$m_{\max}^{1/\beta}$ as function of $\eta$ for 
different system sizes and for a non-additivity 
$\Delta = 2$ on Fig.~\ref{mean2}a.
Second, we plot the rescaled average order parameter 
$\langle |m| \rangle^{1/\beta}$ for the same system 
sizes but for a non-additivity $\Delta = 1$ on 
Fig.~\ref{mean2}b. The linear behavior 
observed on both figures for a critical exponent 
$\beta = 1/8$ confirms the Ising universality class 
prediction. Similar results are obtained for all 
the non-additivities and for the WR model. 
The range of the linear regime is still rather small 
and limited for large packing fractions to 
$m_{\max}^{1/\beta} < 0.7$ and $\langle |m| 
\rangle^{1/\beta} < 0.6$. Furthermore, due to the 
strictly positive average of the absolute order 
parameter, the power law behavior is also limited 
from below to $\langle |m| \rangle^{1/\beta} > 0.1$.
The restriction on the range of the power law behavior
prevents from  a direct determination of the critical 
exponent $\beta$. 

\begin{figure}
\includegraphics[width=8cm]{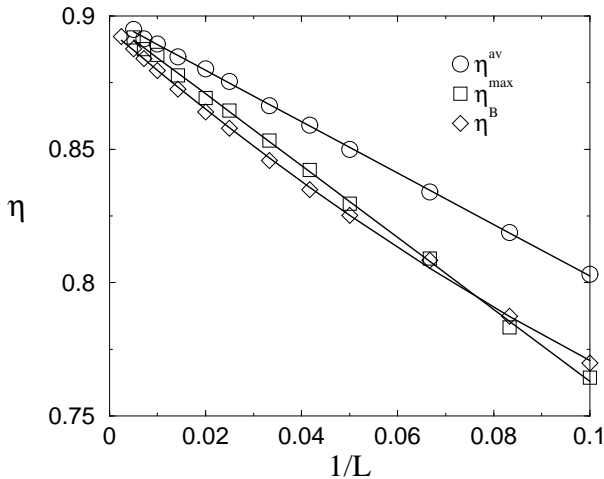}
\caption{\label{dens1}
Finite size critical packing fractions $\eta^{av}_c(L)$, 
$\eta^{\max}_c(L)$ and $\eta^{B}_c(L)$ as function of 
the rescaled system size $1/L^{1/\nu}$.
The non-additivity considered is $\Delta = 1$. 
Linear fits of $\eta^{av}_c$ and $\eta^{\max}_c$ 
confirm the critical exponent $\nu = 1$ of the 
Ising universality class and allow us to extract 
the critical packing fractions $\eta_c^{av}$ and 
$\eta_c^{\max}$. A quadratic fit is necessary for 
$\eta^{B}_c(L)$ in order to extract the critical 
packing fraction $\eta^{B}_c$ due to the corrections 
to scaling and stronger finite size effects.}
\end{figure}

The critical packing fraction $\eta_c$ is then 
determined from the plot of the finite size critical
packing fraction $\eta_c(L)$ as function of the 
rescaled system size $1/L^{1/\nu}$ with the expected
exponent $\nu = 1$. On Fig.~\ref{dens1}, the 
numerical results for $\eta_c^{\max}$, $\eta_c^{av}$ 
and $\eta_c^{B}$ are plotted for a non-additivity 
$\Delta = 1$. The results for $\eta_c^{\chi}$ 
close to those for $\eta_c^{B}$ have been removed
for clarity. The error bars are smaller than the 
symbols and thus not represented on the figure.
As can be seen on Fig.~\ref{dens1}, $\eta_c^{\max}$
and $\eta_c^{av}$ present a linear behavior with
respect to $1/L^{1/\nu}$ for $\nu = 1$. This confirms
the value of the critical exponent $\nu$ predicted
by the Ising universality class. However, 
$\eta_c^{B}(L)$ presents corrections to scaling 
and a quadratic fit is necessary to extract the
critical packing fraction $\eta_c^B$.  
The same is true for $\eta_c^{\chi}$ and similar
results are obtained for all the non-additivities 
as well as the WR model. 

\begin{table}
\begin{tabular}{|c|c|c|c|c|}
\hline
$\Delta$ & $\eta_c^{av}$ & $\eta_c^{\max}$ & $\eta_c^{B}$  & 
$\eta_c^{\chi}$ \\
\hline
\ $0.5$ \ & $0.8141(5)$ & $0.8138(11)$ & \ $0.811(2)$ \ &
\ $0.811(4)$ \ \\ 
\hline
$1.0$ & $0.8988(8)$ & $0.8976(20)$ & $0.896(4)$ & $0.895(5)$ \\
\hline
$2.0$ & \ $1.0215(18)$ \ & \ $1.0179(45)$ \ & $1.017(9)$ 
& $1.017(6)$ \\
\hline
$4.0$ & $1.147(4)$ & $1.141(5)$ & $1.141(5)$ & $1.140(5)$ \\
\hline
$\infty$ & $1.228(3)$ & $1.222(6)$ & $1.224(8)$ & $1.225(5)$ \\
\hline
\end{tabular}
\caption{\label{density2}
Numerical results for the critical packing fraction 
for the two dimensional NAHC mixtures for different 
non-additivities and for the WR model ($\Delta = \infty$). 
The numbers in parenthesis correspond to the error 
on the last digits.}
\end{table}

All the critical packing fractions from the 
four different definitions of their finite 
size analog are presented on Table~\ref{density2} 
for all the non-additivities considered and for 
the WR model. It is interesting to notice that 
the four different definitions are not identical 
since the numerical values obtained for finite 
sizes differ (see Fig.~\ref{dens1}). However, 
the thermodynamic limit results for the critical 
packing fractions are consistent for all the four 
definitions. The estimation of the relative error
which is smaller than $1\%$ for all the critical
packing fractions combines the error on the finite 
size estimates and the error coming from the 
finite size scaling to extract the thermodynamic 
limit results. 

Let us now compare our results to previous
simulations. Our prediction for the critical 
density $\rho_c \sigma_{AB}^2 = 1.560(10)$ 
of the WR model is in good agreement with 
the value $1.566(3)$ obtained by Johnson et 
al.~\cite{Johnson}. For the NAHC mixtures 
in two dimensions, the determination of the 
critical packing fractions have been recently 
obtained from numerical simulations by Saija 
and Giaquinta~\cite{Saija2}.
Only small non-addtivities $\Delta \leq 1$ have 
been considered and the system was limited to 
$800$ particles. The critical packing fraction
was determined without finite size scaling and 
may thus be considered only as a lower bound.
Their critical packing fraction obtained for 
$\Delta = 1$ is $\eta_c = 0.118 (1+\Delta)^2 
= 0.472$. A scaled particle theory predicts 
$\eta_c = 0.096 (1+\Delta)^2 = 0.385$~\cite{Mazo}
whereas a virial expansion $0.324$. 
The different definitions of the packing fraction 
in both cases explains the $\Delta$ dependence 
introduced to compare to our prediction $\eta_c = 
0.897(2)$. The strong difference should be 
contrasted by the fact that the prediction of a 
first-order perturbation theory usually referred
to as MIX1 gives higher values for the critical 
packing fractions especially for large 
non-additivities ($\Delta \geq 0.4$)~\cite{Saija2}.
It could be interesting to pursue the comparison 
of this theory for higher non-additivities 
($\Delta > 1$) with our numerical predictions. 

\subsection{Three dimensional NAHC mixtures}

In three dimensions, we considered a large 
number of different non-additivities $\Delta 
= 0.5$, $0.6$, $0.8$, $1.0$, $2.0$, $3.0$,
$4.0$ and $9.0$ as well as the WR model 
($\Delta = \infty$). The system sizes ranged 
from $L = N_A^{1/3} = 5$ to $50$. The largest 
systems contained $N_A + N_B = 250 000$ particles. 
Systems with up to $2 \times 10^6$ (or $L = 100$) 
where simulated for some non-additivities. The 
equilibrium time was then close to the day on 
simple personal computers. Since a large number 
of densities needs to be simulated to extract the 
critical density, we are for $L = 100$ at the 
limit of the numerical capabilities. As for the 
two dimensional systems, five consecutive runs 
were simulated. The first one was used to 
equilibrate the initial configuration and the 
four remaining ones served for the collection 
of data and the estimation of errors. 
$10^5$ to $10^6$ Monte-Carlo steps for each
run were sufficient for the equilibration and 
in order to obtain a small error.

\begin{figure}
\includegraphics[width=8cm]{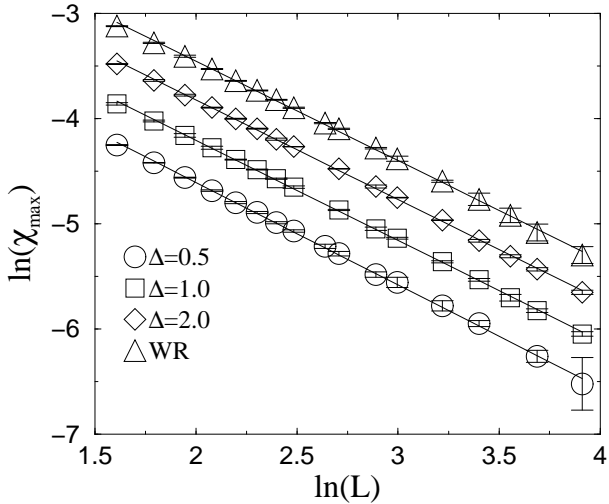}
\caption{\label{susmax3}
Maximum of the modified susceptibility $\chi_{\max}$ 
as function of the system size $L$ in a log-log scale
for different non-additivities $\Delta$ and for the
WR model (Widom). The ratio of critical exponents 
$\gamma/\nu$ is extracted from linear fits. Data have 
been shifted for clarity.}
\end{figure}

The modified susceptibility $\chi$ is determined 
as function of the packing fraction $\eta$ for 
different system sizes. It presents a single 
maximum $\chi_{\max}(\eta_c^{\chi}(L),L)$ at the 
finite size critical packing fraction 
$\eta_c^{\chi}(L)$. As can be seen on 
Fig.~\ref{susmax3}, this maximum depends on the 
system size with a power law behavior. From a 
linear fit of the maximum of the modified
susceptibility as function of the system size 
in a log-log scale, we deduce the ratio of critical 
exponents $\gamma/\nu$ (see Table~\ref{gammanu}). 
Those ratios for all non-additivities are 
systematically larger than the Ising universality 
class prediction but still inside the error bars. 
The relative errors were estimated as the sum of 
the average relative error on the maximum $\chi_{\max}$ 
and the error due to the linear fit.

The existence of corrections to scaling may 
explain the over-estimation of $\gamma/\nu$. 
Removing the data from the small size systems 
for the linear fits reduces the discrepancy 
with the Ising universality class prediction. 
A possible source for the corrections to 
scaling observed is the existence of 
fluctuations on the density inside each box 
during the simulations. Even though those 
fluctuations are small they may affect 
slightly the scaling behavior especially 
for small systems were the fluctuations are 
stronger. Some constrained simulations have 
been done were the density in each box was 
kept constant. The clusters were moved only 
if they did not change the total number of 
particles inside each box. 
This modification of the algorithm leads to
similar results with less pronounced corrections 
to scaling. However, errors were also stronger 
since due to the rejections the equilibration 
time was sensibly increased. 

\begin{figure}
\includegraphics[width=8cm]{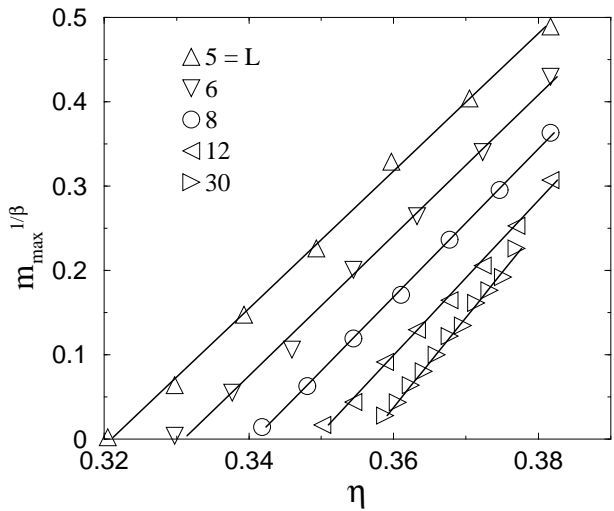}
\caption{\label{xmax3}
Rescaled maximum of the distribution 
of the order parameter $m_{\max}^{1/\beta}$ 
as function of the packing fraction $\eta$. 
The non-additivity considered is $\Delta = 0.8$. 
The critical exponent $\beta = 0.326$ predicted 
for the Ising university class has been used
and the critical packing fractions $\eta_c^{\max}
(L)$ for different system sizes $L$ are extracted 
from linear fits.}
\end{figure}

The confirmation of the critical exponent $\beta$
is obtained from the plot of the rescaled 
maximum of the distribution of the order parameter 
$m_{\max}^{1/\beta}$ as function of the packing 
fraction $\eta$ for different system sizes.  
A non-additivity $\Delta = 0.8$ has been considered
on Fig.~\ref{xmax3} but similar results are obtained 
for all non-additivities and for the WR 
model. The linear behavior observed on 
the figure for a critical exponent $\beta = 0.326$ 
confirms the Ising universality class prediction. 
The range of the linear regime is still rather 
small and limited for large packing fractions to 
$m_{\max}^{1/\beta} < 0.5$. The range of the linear 
regime for $\langle |m| \rangle^{1/\beta}$ is even 
smaller and cannot allow us the confirmation of the 
critical exponent $\beta$ neither the determination 
of the finite size critical packing fraction 
$\eta_c^{av}(L)$.

\begin{figure}
\includegraphics[width=8cm]{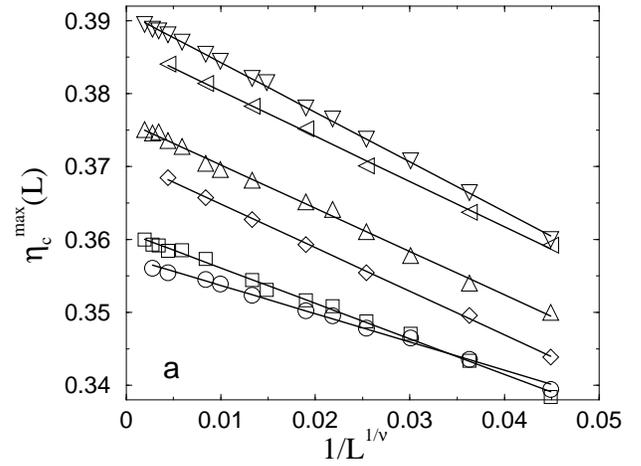}
\includegraphics[width=8cm]{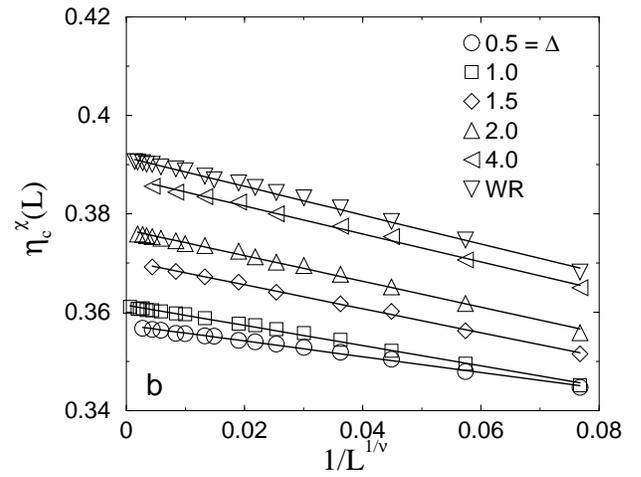}
\caption{\label{etasuscep}
Finite size critical packing fractions 
$\eta_c^{\max}(L)$ (a) and $\eta^{\chi}_c(L)$ 
(b) as function of the rescaled system 
size $1/L^{1/\nu}$ with $\nu = 0.627$ 
as predicted by the Ising universality class.  
Different non-additivities are considered as
well as the WR model (same legends for both 
figures). Linear fits of $\eta_c^{\max}$ as 
well as $\eta^{\chi}_c$ confirm the critical 
exponent $\nu = 0.627$ and allow us to extract 
the critical packing fractions $\eta_c^{\max}$ 
and $\eta_c^{\chi}$. }
\end{figure}

The critical packing fraction $\eta_c$ is  
determined from the plot of the finite size 
critical packing fraction $\eta_c(L)$ as 
function of the rescaled system size $1/L^{1/\nu}$ 
with the expected exponent $\nu = 0.627$. 
The numerical results for $\eta_c^{\max}(L)$ 
(Fig.~\ref{etasuscep}a) and for $\eta^{\chi}_c(L)$ 
(Fig.~\ref{etasuscep}b) are plotted for different 
non-additivities and for the WR model. The error bars 
are smaller than the symbols and thus not represented. 
As can be seen, $\eta_c^{\max}$ and $\eta_c^{\chi}$ 
present a linear behavior with respect to $1/L^{1/\nu}$ 
for $\nu = 0.627$. This confirms the value of the 
critical exponent $\nu$ predicted by the Ising 
universality class. The results are similar for 
$\eta_c^B$ and for the other non-additivities 
considered. The corrections to scaling observed
for $\eta_c^{\chi}$ and $\eta_c^{B}$ in two 
dimensions are not present in the three dimensional 
systems. It is possible to notice a change in the 
slope of the linear fit around $\Delta = 1$. 
Already evident on Fig.~\ref{etasuscep}, this is 
confirmed by the results for the non-additivities 
$\Delta = 0.6$ and $0.8$ (data not shown on the figure).
For $\Delta > 1$ the slope is independent of the 
non-additivity whereas for $\Delta < 1$ the slope 
is increasing with the non-additivity. This change
may be due to the onset of the fluid-solid transition
that the model experience for large packing fractions.
At sufficiently low non-additivity, the phase 
separation transition is expected to be pre-empted 
by the fluid-solid transition~\cite{Mazo}. 

\begin{table}
\begin{tabular}{|c|c|c|c|}
\hline
$\Delta$ & $\eta_c^{\max}$ & $\eta_c^B$ & $\eta_c^{\chi}$\\
\hline
\ $0.5$ \ & $0.3577(3)$ & $0.3574(3)$ & $0.3574(3)$\\
\hline
$0.6$ & $0.3572(4)$ & $0.3564(4)$ & $0.3568(4)$\\
\hline
$0.8$ & $0.3583(5)$ & $0.3581(4)$ & $0.3580(5)$\\
\hline
$1.0$ & $0.3610(5)$ & $0.3614(3)$ & $0.3614(5)$\\
\hline
$1.5$ & $0.3706(5)$ & $0.3703(5)$ & $0.3705(8)$\\
\hline
$2.0$ & $0.3758(5)$ & $0.3766(5)$ & $0.3766(8)$\\
\hline
$3.0$ & $0.3827(10)$ & $0.3839(6)$ & $0.3843(9)$\\
\hline
$4.0$ & \ $0.3862(10)$ \ & \  $0.3871(8)$ \  & 
\ $0.3872(11)$ \ \\
\hline
$9.0$ & $0.3896(9)$ & $0.3908(6)$ & $0.3914(9)$ \\
\hline
$\infty$ & $0.3910(4)$ & $0.3912(4)$ & $0.3912(4)$\\
\hline
\end{tabular}
\caption{\label{density3}
Numerical results for the critical packing 
fractions in three dimensions for the 
different non-additivities considered and 
for the WR model ($\Delta = \infty$). 
The numbers in parenthesis correspond to 
the error on the last digits.}
\end{table}

All the critical packing fractions for the 
different non-additivities and for the 
WR model are presented on Table~\ref{density3}. 
The relative errors are less than $0.5\%$ and 
the three definitions of the critical packing 
fraction $\eta_c^{\max}$, $\eta_c^{\chi}$ and 
$\eta_c^{B}$ lead to identical estimations
in the thermodynamic limit. 

Let us now compare our results with
others. Our prediction for the critical density 
of the WR model is $\rho_c \sigma_{AB}^3 = 0.7470(8)$ 
which is in strong agreement with the value $0.748(2)$ 
obtained by Johnson et al.~\cite{Johnson} with 
another cluster algorithm. Our result improves by 
a factor two the error on this critical density.
The critical packing fraction has been determined
from different numerical simulations for a large
set of non-additivities. For a non-additivity 
$\Delta = 1$, our prediction $\eta_c/(1+\Delta)^3 
= 0.04515(8)$ is slightly higher than the value 
$0.04484(20)$ proposed by G\'o\'zd\'z~\cite{Gozdz}. 
However, it strongly confirms the previous 
under-estimation of $0.0288$ by Saija et 
al.~\cite{Saija}. The non-additivity dependence 
introduced is due to a different definition for 
the packing fraction. G\'o\'zd\'z~\cite{Gozdz} 
also predicted the critical packing fraction 
$0.0866(2)$ and $0.0611(2)$ for the non-additivity 
$\Delta = 0.6$ and $0.8$ respectively. Its predictions 
compare nicely with our results $\eta_c/(1+\Delta)^3 = 
0.0871(1)$ and $0.0614(1)$. It is important to notice 
that the critical packing fraction is determined to our 
knowledge for non-additivities $\Delta > 1$ for the 
first time in this study. This is possible thanks
to the absence of a critical slowing down of 
the simulations for large non-additivities 
with the cluster algorithm in comparison to 
the simulations in the semigrand canonical 
ensemble. 

\begin{figure}
\includegraphics[width=8cm]{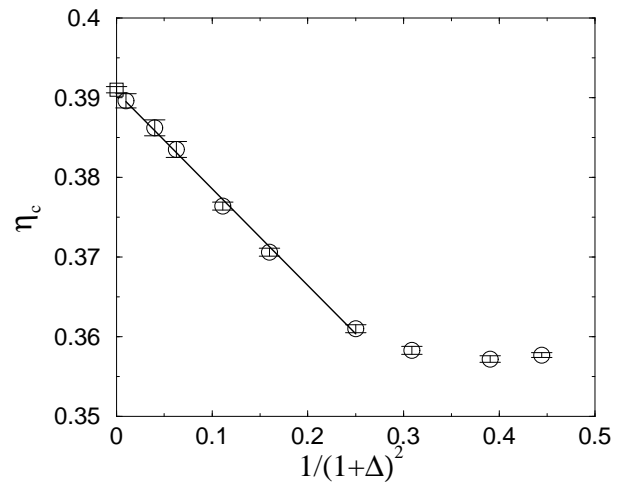}
\caption{\label{density}
Critical packing fraction for the different 
non-addtivities $\eta_c(\Delta)$ as function 
of $1/(1+\Delta)^2$ (disks) and for the 
WR model (square). The line 
corresponds to a linear fit for the six 
largest non-additivities and leads to an 
estimation of $\eta_c = 0.3907(5)$ for 
the WR model.}
\end{figure}

From the determination of the critical packing
fraction for large non-additivities, it is 
possible to consider the convergence of the 
NAHC mixtures to the WR model which 
corresponds to an infinite non-additivity. 
It is evident from Fig.~\ref{density} that the 
critical packing fraction $\eta_c (\Delta) 
= \eta_c(\infty) - B/(1+\Delta)^2$ for large
non-additivities. A linear fit for the six 
largest non-additivities leads to an estimate 
of $\eta_c = 0.3907(5)$ for the WR model in 
good agreement with the direct simulations. 
The coefficient $B$ was estimated to be $1.21$.
This interesting dependence of $\eta_c(\Delta)$ 
on $\Delta$ may be a good test for the different 
theories developed to determine the critical 
packing fraction of the NAHC mixtures.

\begin{figure}
\includegraphics[width=8cm]{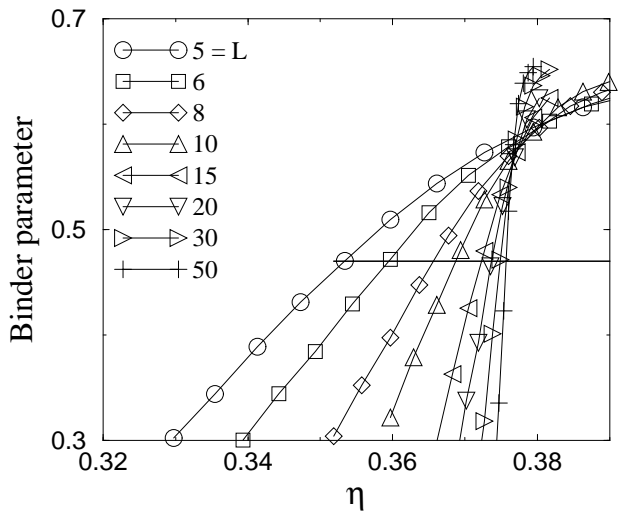}
\caption{\label{Binder3}
Binder parameter for different system sizes
as function of the packing fraction $\eta$ 
for the non-additivity $\Delta = 2.0$. The 
horizontal line represents the universal 
intersection value $U^* = 0.47$ expected 
for the Ising universality class.}
\end{figure}

The Binder parameter offers another test of the 
universality class from its intersection value 
$U^*$. In Fig.~\ref{Binder3}, we have plotted 
the Binder parameter as function of the packing 
fraction for different system sizes and for the 
non-additivity $\Delta = 2$. It can be seen that 
the intersections between the different curves 
occur at smaller and smaller values as the system 
size increases but still at higher values than 
the expected universal one $U^* = 0.47$. Similar 
results are observed for the other non-additivities 
and for the WR model. This result was already 
observed in G\'o\'zd\'z simulations~\cite{Gozdz}. 
Corrections to scaling may explain such behavior.
For our definition of the critical packing fraction 
$\eta_c^B$ based on the Binder parameter, the 
arbitrary choice was $U(\eta_c^{B}(L),L) = 1/2 > 
U^*$. Thus, $\eta_c^B(L)$ was expected to decrease 
with increasing system size. The opposite behavior
is found. The absence of corrections to scaling 
in the dependence of $\eta_c^B(L)$ with the system
size is thus an argument against this explanation
for the high intersection in Fig.~\ref{Binder3}.
However, no other explanations have been found. 

\section{Discussion}

In this paper we present a cluster algorithm
for the numerical simulations of the NAHC 
mixtures as well as for the WR model. 
This cluster algorithm allows one to simulate
large systems (up to $10^6$ particles). 
Each Monte-Carlo step corresponds to the 
non-local move of a large number of particles 
reducing strongly the equilibration time.
The absence of critical slowing down
for increasing non-additivities $\Delta$
allows us to study large non-additivities 
and the convergence of the NAHC mixtures 
to the WR model when $\Delta \rightarrow 
\infty$. 

Two and three dimensional systems have been 
considered. In both cases, the models were 
found to belong to the Ising universality 
class for all the non-additivities considered
as well as for the WR model. The critical 
packing fractions $\eta_c$ have been 
determined from four different finite 
size definitions. All the definitions lead 
to identical predictions in the thermodynamical
limit with a high precision. 

Different theories have been used to 
determine the critical packing fractions 
$\eta_c(\Delta)$ of NAHC mixtures~\cite{Santos}, 
from the scaled particle theory~\cite{Mazo} to 
the virial expansion~\cite{Saija3,Saija4} 
or the density functional theory~\cite{Schmidt}.
A comparison with our precise numerical 
predictions of the critical packing fractions 
for a large set of non-additivities $\Delta$
is possible. Even if a quantitative comparison 
between theory and simulations is difficult, 
the $\Delta$ behaviour of the critical 
packing fraction ($\eta_c = \eta_c(\infty) - 
B/(1+\Delta)^2$ in three dimensions) for large 
non-additivities $\Delta$ is a good test for a 
theory.

Recently, Jagannathan and Yethiraj~\cite{Jagannathan2}
have studied the dynamical behavior of the 
WR model close to the phase separation 
transition. The cluster algorithm could
be used to consider larger systems in order
to be closer to the critical density.
The cluster algorithm could also be used 
to study the stability and interfacial 
properties in confined geometries~\cite{Duda}.

\subsection*{Acknowledgments}

W. Krauth is thanked for useful discussions and a
careful reading of the manuscript.

\end{document}